# Hyperspectral Absorption Microscopy Using Photoacoustic Remote Sensing

Kevan Bell[1,2], Lyazzat Mukhangaliyeva[1], Layla Khalili[1], Parsin Haji Reza[1,*]

[1]*PhotoMedinine Labs, Department of Systems Design Engineering, University of Waterloo, Waterloo, Ontario N2L 3G1, Canada*
[2]*illumiSonics Inc., 22 King Street South, Suite 300, Waterloo, Ontario N2J 1N8, Canada*
*\*Corresponding Author: phajireza@uwaterloo.ca*

**Abstract:** An improved method of remote optical absorption spectroscopy and hyperspectral optical absorption imaging is described which takes advantage of the photoacoustic remote sensing detection architecture. A wide range of photoacoustic excitation wavelengths ranging from 210 nm to 1550 nm was provided by a nanosecond tunable source allowing access to various salient endogenous chromophores such as DNA, hemeproteins, and lipids. Sensitivity of the device was demonstrated by characterizing the infrared absorption spectrum of water. Meanwhile, the efficacy of the technique was explored by recovering cell nuclei and oxygen saturation from a live chicken embryo model and by recovering adipocytes from freshly resected murine adipose tissue. This represents a continued investigation into the characteristics of the hyperspectral photoacoustic remote sensing technique which may represent an effective means of non-destructive endogenous contrast characterization and visualization.

## Introduction

Hyperspectral imaging is rapidly becoming an essential tool for investigating the world around us. By effectively providing the chromophore specificity present in a spectrometer along with the spatial discrimination afforded by an imaging technique, these devices generate a wealth of information useful across fields and disciplines. Conventional hyperspectral imaging via light transmission, reflection, and fluorescence can be used across a wide range of clinical, pre-clinical, and nondestructive testing pursuits. These techniques have demonstrated efficacy from everything to predicting crop health [1] or monitoring of food spoilage [2], to providing noninvasive disease diagnosis and surgical guidance [3]. However, there remains an unmet need for one such all-optical approach for hyperspectral imaging that can provide direct optical absorption measurement in reflection mode. This capability would be highly valuable given the extensive range of endogenous optical absorption contrast available within biological tissues.

A potential solution to this conundrum comes from the field of photoacoustic modalities. This family of optical techniques excite thermo-elastic pressure waves within a sample using a short (commonly nanosecond) excitation pulse. As these pulses will be absorbed by specific chromophores, the generated acoustic profiles encode spatial information regarding the distribution of said chromophores. Lower-resolution high-penetration embodiments of this technique commonly utilize a tunable flash lamp driven optical parametric oscillator (OPO) providing high pulse energies (mJ) at mediocre repetition rates (<50 Hz) appropriate for use in photoacoustic tomography (PAT) [4]. However, such optical sources become inappropriate when moving to higher-resolution laser-scanning-based devices such as optical-resolution photoacoustic microscopy (OR-PAM) [5] or photoacoustic remote sensing (PARS®) microscopy [6, 7]. Since these devices scan point-by-point, they require higher repetition rates, and demand high beam quality to achieve near-diffraction-limited lateral performance on the sample. This has commonly led to the implementation of white-light optical sources [8] which may not provide sufficient optical fluence at the sample when a narrow-line excitation is desired. Moreover, for some sensitive samples such as open wounds or open surgical sites, the use of contact-based approaches such as OR-PAM may be impractical or incompatible with the desired target.

Here, we present a hyperspectral optical absorption imaging microscope based on a PARS detection pathway. The PARS approach replaces the acoustic transducer common to conventional OR-PAM devices with a co-focused detection beam that encodes photoacoustic thermal and pressure effects as back-reflected intensity perturbations. This non-contact photoacoustic device features a high-speed OPO driven by a nanosecond diode-pumped solid-state laser configured to deliver pulses at up to 1 kHz in the range from 210 nm to 1550 nm. The tunable nature of this device provides access to a wealth of endogenous biological contrast and structure. Exciting in ultraviolet (UV) (~260 nm) targets DNA contrast showing subcellular nuclear structural information. Exciting in the visible range (~400 nm to ~600 nm) targets hemeproteins such as cytochromes providing cytoplasmic structure, and hemoglobin highlighting erythrocytes and blood vessel structure. Finally, exciting in the infrared (IR) (~1200 nm) targets lipid and collagen. This is of course not an exhaustive list.

In this article we explore the efficacy of this improved device operating at salient wavelengths for various specimen. We have previously reported works related to an earlier embodiment of this system looking at formalin-fixed paraffin-embedded (FFPE) human tissue slides, FFPE human tissue blocks, frozen sections of human skin taken from Mohs micrographic surgery (MMS), and freshly excised unprocessed murine organs [7]. In these works, primarily DNA (250 nm) and hemeproteins (420 nm) were investigated highlighting cellular structure, erythrocytes, and cytoplasm. Moreover, the capabilities of the previous device were demonstrated operating as both a non-contact spectrometer and a hyperspectral absorption microscope in simple chromophore solutions and FFPE targets [9]. Here, this second-generation system with improved wavelength range was used to investigate live specimens by imaging chicken embryo chorioallantoic membrane (CAM) to explore *in vivo* DNA, hemoglobin, and lipid contrast for the first time in a single system. As well, unprocessed freshly resected murine adipose tissue was used to investigate lipid-rich samples with the newly implemented infrared excitation arm marking the first such report of a non-contact photoacoustic modality that is capable of visualizing individual adipocytes. These new capabilities are explored and discussed.

## Methods

The HS-PARS system explored in these investigations is similar to that previously reported in [9] with several key differences. (i) The available wavelength range has been extended from 210 nm to 680 nm by adding an additional infrared band from 1050 nm to 1550 nm. This allows access to commonly reported lipid absorption regions (~1210 nm) along with paraffin and water absorption. (ii) The previous reflective objective lens was replaced by a right-angle parabolic mirror. This offers superior power transmission characteristics and improved Gaussian beam propagation both of which are due to the lack of the previous central occlusion. (iii) Finally, the mechanical scanning support system required redesign to accommodate tighter mechanical clearances which arose due to the effective shorter working distance provided by the parabolic mirror.

A diagram of the apparatus is shown in Fig. 1. The system consisted of a tunable excitation provided by a 1 kHz OPO driven by a diode-pumped solid-state laser (NT242, Ekspla). This output was attenuated and passed through a series of dichroic mirrors to separate four wavelength bands (210 nm to 405 nm (UV); 405 nm to 505 nm (blue); 505 nm to 680 nm (green-red); and 1050 nm to 1550 nm (IR)). This was done such that we could manage each sub-range with appropriate achromatic optic systems and spatial filters. Spatial filters were required due to the highly astigmatic and elliptical output from the OPO. Three of these spatial filters (UV, blue, and IR) were implemented using a pair of air-spaced doublets surrounding a pinhole. Meanwhile the green-red pathway spatial filter utilized a single mode optical fiber. These four excitation bands passed through variable beam expanders (UV, blue, and IR) and a variable collimator (green-red) to provide variable divergence allowing for compensation of chromatic effects at the sample. Splitting the source range also allowed for normalization of pulse energies which varied significantly across the wavelength range. Another series of dichroic mirrors was used to recombine the excitation pathways along with the detection beam and direct them to the sample. Detection was provided by a 970 nm superluminescent diode (SLD970P-A40W, ThorLabs). This fiber-coupled source was collimated using a variable collimator and passed through a polarizing beamsplitter (PBS) and quarter-waveplate (QWP) before joining the excitation pathways in the dichroic stack. Back-reflected detection from the sample passed back through the QWP and PBS being redirected through a bandpass filter and being focused on to a photodiode using an aspheric condenser lens. Mechanical scanning of the sample about the optical interrogation point was provided by two linear stepper motor stages. The system could be used as both an absorption spectrometer by scanning a given region over fine excitation wavelength increments or could be used as a hyperspectral imaging platform by imaging at several salient wavelengths.

Several sample types were explored with this system including *in vivo* CAM from chicken embryos and freshly resected murine tissues. The CAM models used in these experiments were developed in house. Fertilized White Leghorn eggs were placed in a hatcher for 72 hours before being cracked and incubated for an additional 10 days in an open vessel. These were then flipped into containers featuring UV-transparent cover slips such that the CAM could be pressed against this window to stabilize it for mechanical scanning. This allows for scanning of the live subject by the system. Freshly excised murine tissues were obtained with the aid of collaborators at the Central Animal Facility, University of Waterloo under protocol ID: 41543 (Photoacoustic Remote Sensing (PARS) Microscopy of Resected Rodent Tissues, University of Waterloo). All murine tissue experiments were performed in accordance with the relevant guidelines and regulations. Freshly excised tissues were immediately placed in phosphate buffer solution and imaging was conducted within 3 hours of devitalization.

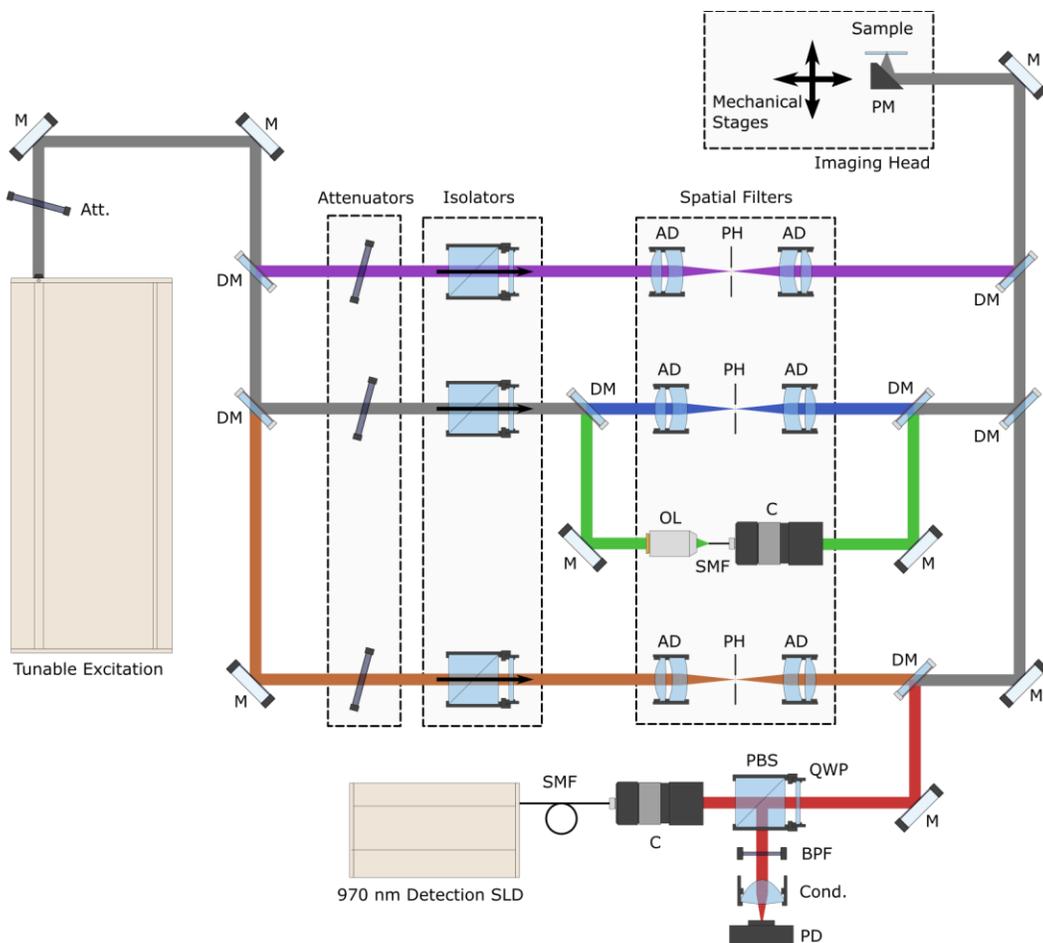

**Fig. 1.** System diagram of the reported system. Shorthand notation is defined: (AD) Achromatic Doublet; (Att.) Attenuator; (BPF) Bandpass Filter; (C) Collimator; (Cond.) Condenser; (DM) Dichroic Mirror; (M) Mirror; (OL) Objective; (PBS) Polarizing Beamsplitter; (PD) Photodiode; (PM) Parabolic Mirror; (QWP) Quarter Waveplate; (SMF) Single-mode Fiber.

## Results & Discussion

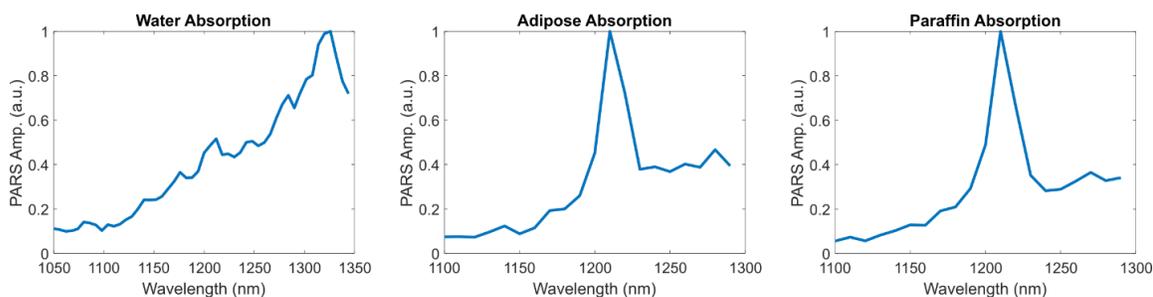

**Fig. 2.** Several absorption measurement sweeps using the infrared wavelength band showing water, FFPE adipose tissue, and paraffin.

The performance of the infrared excitation band was examined first. Three chromophores of interest in this range include water, adipose tissue, and paraffin. These sweeps are shown in Fig. 2. All three of these sweeps agree well with previously reported trends [10-12]. Measurements suggest that the HS-PARS system should be able to distinguish lipid-rich tissues from the background water signal by targeting the distinct 1210 nm peak. However, the lipid and paraffin signals overlapped significantly within this spectral region, likely preventing distinct recovery of adipocytes in paraffin-embedded samples. This was later confirmed in experiments with formalin-fixed paraffin-embedded (FFPE) blocks and slides where no distinct lipid contrast was observed at 1210 nm or other nearby wavelengths. Moreover, experiments involving formalin-fixed tissue chunks also yielded poor infrared contrast suggesting crosslinking from formalin fixation may significantly alter available lipid contrast in these samples. Water measurements were conducted by placing a small drop of distilled water onto a UV-transparent cover slip with the system targeting the air-water boundary to provide maximum signal amplification. Measurements in FFPE blocks targeted the air-block interface for the same purpose.

The hyperspectral imaging system next explored available contrast from chicken embryo CAM. Acquisitions were performed using 300 nm lateral step sizes. One such dataset is presented in Fig. 3. First, ultraviolet 250 nm is used to excite DNA contrast Fig. 3a. The wavelength was selected as previous studies [7] have found 250 nm to provide optimal contrast between cell nuclei and the background tissue despite the slight displacement from the ~260 nm DNA absorption peak. These results represent the first such report of label-free visualization of live cell nuclei in a CAM model. Moreover, multiple distinct cell layers can be seen in this image with several more yet slightly out of focus of the system. This is consistent with the CAM anatomy which features two relatively cell-dense epithelium layers surrounding the sparser intermediate mesodermal internal layer where larger blood vessels are located. Next, several visible wavelengths were selected to target hemeprotein absorption (420 nm, 532 nm, and 560 nm; Fig. 3b-d respectively) with hemoglobin likely being the dominant such chromophore in these instances. These acquisitions highlighted the fine capillary beds of the CAM. Discrepancies between these three wavelengths likely result from sample motion as each wavelength acquisition required ~16 min to acquire, and the live chicken embryo is free to move during imaging.

Consistency between the 532 nm and 560 nm acquisitions was sufficient to extract blood oxygen saturation (sO$_2$). Unmixing was performed assuming extracted PARS signals represented linear super positions of constituent chromophores (oxy- and deoxyhemoglobin in this case) while accounting for wavelength-dependent excitation fluence. In addition, absorption of the detection wavelength was also considered in this model as it too will vary depending on chromophore concentration resulting in an inverse dependence on the collected PARS signals. The extracted sO$_2$ is shown in overlay along with the DNA contrast in Fig. 3e. One aspect to consider between different wavelengths and target structures is that of lateral resolution. In the case of targeting cell nuclei with ultraviolet, high lateral resolution (<500 nm) is desirable to elicit clear cellular and sub-cellular morphology. However, when trying to visualize the capillary beds such high resolution recovers only a sparse collection of erythrocyte-scale (~6 µm) structures. This is likely attributed to the true sparse nature of these cells within the capillaries where they will be oriented in a single file separated by plasma. As such, visualizations of more connected capillary structures required reduction of lateral resolution (>~4 µm) which provides more consistent and connected structures. This reduction in lateral resolution was accomplished by narrowing the visible wavelength beams relative to the UV beams before entering the common objective. Near-infrared wavelengths were also used on the subjects to target the 1210 nm lipid absorption peak, however, no signal could be recovered from these excitation wavelengths from the CAM. This could be due to several factors. We do not expect significant lipid contrast as this tissue does not feature adipocytes, and the lipid bilayers of cells within this region may not constitute sufficient contrast for the current sensitivity and resolution of the system. Moreover, water absorption may be expected for this target, however, intertissue-interfaces may not provide sufficient scattering contrast to amplify water signals above the noise floor of the system. Standalone water measurements utilized an air-water interface to amplify PARS signals. Meanwhile, tissue-tissue interfaces may provide roughly two orders of magnitude or less scattering contrast as compared to an air-water interface.

To explore and characterize the lipid visualization capabilities of the system, freshly resected murine adipose tissue was imaged using 1210 nm excitation along with 250 nm excitation to highlight DNA contrast (Fig. 4). These captures were taking using 900 nm lateral step sizes. Acquisition of lipid contrast required substantially higher pulse energies as compared to targeting hemoglobin in the live subjects. This is likely attributed to the significant reduction in absorption coefficient. At the local absorption peak of 1210 nm, lipid is expected to provide around two orders of magnitude less absorption as compared to hemoglobin at 532 nm. This, when combined with the larger focal spot provided by the infrared wavelengths as compared to their visible counterparts, demanded substantial pulse energies nearing single micro-Joule levels. As well, this contrast appeared to generally give more vague structure as compared to the well-defined cell nuclei and micro vessel structures seen from the CAM models. This is in part likely do to an expected overall reduction in resolution as a result of the longer wavelengths in use along with targeting larger amorphous structures (adipocytes). However, adipocyte-scale structures (~30 – 50 µm) were recovered in these samples. This represents the first such report of a non-contact photoacoustic modality which can recover lipid contrast, and one of the first photoacoustic microscopes capable of cleanly resolving individual adipocytes.

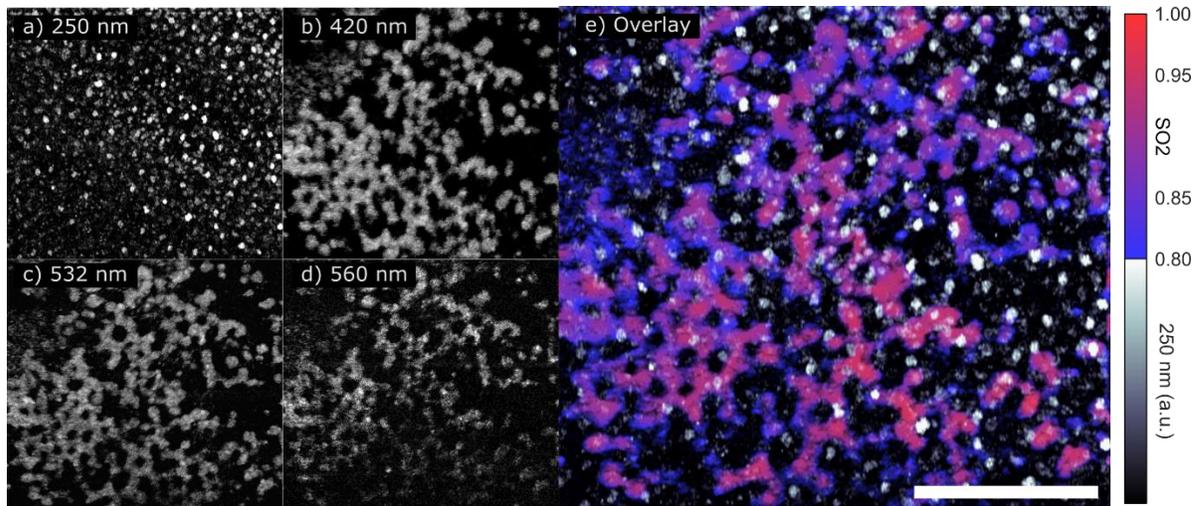

**Fig. 3.** Results from several wavelengths looking at a single region of a chicken embryo CAM. (a-d) Shows each individual acquisition at the respective wavelengths. (e) Shows a processed overlay of the UV acquisition highlighting DNA along with extracted $sO_2$ values. Scale bar: 100 µm.

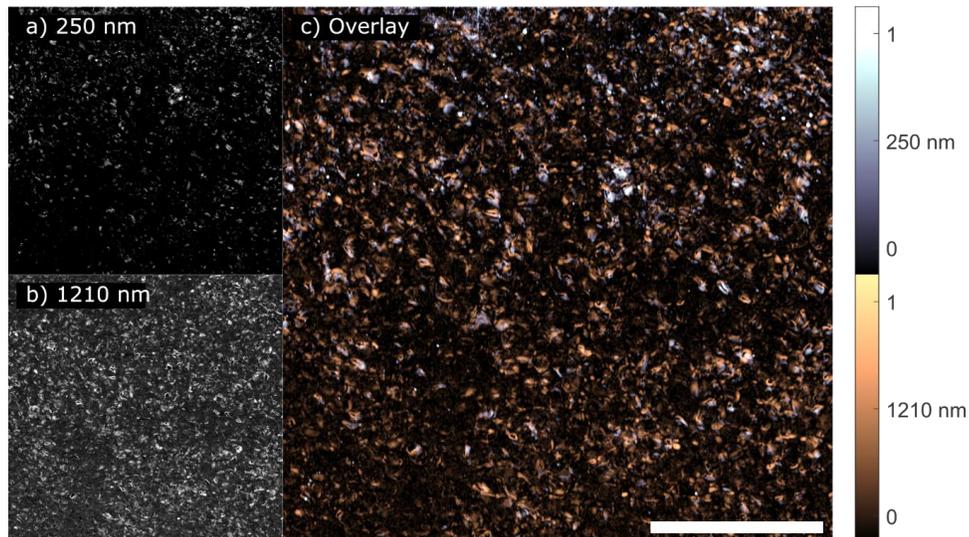

**Fig. 4.** HS-PARS image of freshly resected murine adipose tissues using 250 nm excitation to target DNA contrast and 1210 nm excitation to target lipid contrast. Scale bar: 500 µm.

## Conclusion

Here we have presented recent work surrounding an improved hyperspectral PARS microscope. This demonstrates the first report of a non-contact photoacoustic microscope that is capable of recovering DNA, hemeprotein, and lipid contrast from a single device. Multiplex acquisition efficacy was demonstrated by recovering functional $sO_2$ measurements from a live subject. Meanwhile, sensitivity of the device was such that it was able to characterize infrared absorption of water. The HS-PARS microscope is poised to represent an essential material investigation tool capable of non-destructive recovery of a substantial range of endogenous contrast. Work will continue towards further improving system sensitivity and consistency while maintaining, or yet further extending, its wide excitation wavelength range.

## Acknowledgments

The authors would like to acknowledge Jean Flanagan at the Central Animal Facility, University of Waterloo for her help in procuring murine samples.

## Funding

Natural Sciences and Engineering Research Council of Canada (DGECR-2019-00143, RGPIN2019-06134); Canada Foundation for Innovation (JELF #38000); Mitacs (IT13594); University of Waterloo (Startup Funds); Centre for Bioengineering and Biotechnology (CBB Seed fund); illumiSonics Inc. (SRA #083181).

## Disclosures

K.B. and P.H.R. have financial interests in illumiSonics Inc.